## SOFTWARE

# Visualization of gene expression information within the context of the mouse anatomy

Andy Taylor[1*], Kenneth M[c]Leod[1], Chris Armit[2], Richard Baldock[2] and Albert Burger[1,2]

**Abstract**

**Background:** The eMouse Atlas of Gene Expression (EMAGE) is an online resource that publishes the results of *in situ* gene expression experiments on the developmental mouse. The resource provides comprehensive search facilities, but few analytical tools or visual mechanisms for navigating the data set. To deal with the missing visual navigation, this paper explores the application of sunburst and icicle visualizations within EMAGE.

**Results:** A prototype solution delivered a simple user interface that helps the user query EMAGE and generate a sunburst/icicle diagram. An evaluation featuring test subjects from the EMAGE staff studied the visualizations and provided a range of suggested improvements. Moreover the evaluation discovered that in addition to providing a visual means of walking through the data, when grouped, the sunburst delivers an interactive overview that assists with analysing sets of related genes.

**Conclusions:** The sunburst and icicle visualizations have been shown to be effective tools for summarising gene expression data. The sunburst with its space saving radial layout was found especially useful for providing an overview of gene families or pathways. Work is ongoing to integrate these visualizations into EMAGE.

## Introduction

Due to high throughput experiments and information technology, there are now many big data biological resources. It is increasingly clear that users require assistance when navigating, searching and understanding these information silos.

Within the commercial world Business Intelligence (BI) [1] tools are employed to help users interpret numerical information. Whilst many of the basic ideas behind BI are transferrable to the biological world, some are not. BI is focused on numerical data, and thus commonly visualises results as bar charts, pie charts and scatterplots. Whilst these visualizations are ideal for representing the number of transactions within a business, they appear less suitable for depicting the relationships between, for example, genes and tissues.

CUBIST [2] was a project that explored Semantic BI (SBI) - the combination of the traditional BI techniques with semantic technologies - in order to provide relationship analysis that complements the numerical analysis provided by traditional BI. As part of this process, CUBIST developed a dashboard (i.e., analytical workbench) with novel BI visualizations, such as sunburst and hasse diagrams [3]. That work inspired the development of extra, more specialised, visualizations for one CUBIST use case: *in situ* hybridisation gene expression for the developmental mouse.

We are not aware of an attempt to visually present gene expression data for the mouse within the context of the anatomy that does not feature a picture of the mouse. We aim to develop prototype visualizations that address this gap in such a way that the chosen mechanisms make sense to the EMAGE user community. This paper discusses the development and impact of these specialised visualizations.

## Background

This paper focuses on one biological resource as its use case. That resource, EMAGE [4], publishes online gene expression information for the developmental mouse.

A gene is a hereditary unit consisting of a sequence of DNA that occupies a specific location on a chromosome and determines a particular characteristic in an organism. A gene is considered active if it is transcribed resulting in one or more RNA products and, following translation, one or more protein products. This phenomenon of transcription and translation is additionally known as gene expression.

*Correspondence: ajt17@hw.ac.uk
[1]Heriot-Watt University, EH14 4AS Edinburgh, UK
Full list of author information is available at the end of the article

The gene expression information is obtained by experimenting on a mouse embryo. Each embryo corresponds to a point in time of the *developmental mouse*: the mouse from conception until birth. The time window is split into 26 distinct periods called Theiler Stages (TS). Each stage has its own anatomy, and corresponding anatomy ontology, called EMAP [5].

Additionally, there is the notion of an *abstract anatomy*. This is a single anatomy (and matching anatomy ontology) in which the 26 Theiler Stages are aggregated by taking their union. The staged EMAP anatomy has a structure called liver in TS16, a second in TS17 and so on until an eleventh in TS26. In contrast, the abstract anatomy has a single liver, which represents the liver across all the stages in which it exists. Each structure has its own unique identifier of the form EMAP:*number*. For example, the heart in TS12 is EMAP:315, and the heart in TS17 is EMAP:2411. The abstract heart has the identifier EMAPA:16105.

The initial version of EMAP documented the anatomy as a tree with each structure being *partOf* another, e.g., the digit is *partOf* the paw. Subsequent extensions have resulted in a directed acyclic graph (DAG), yet *partOf* is still the predominant relationship (the other relationship is *is a*; however, it is used infrequently). The original tree representation still exists, as a subset of the DAG, and is the favoured visual representation of the anatomy (it only uses *part of*).

The result of an *in situ* hybridization (ISH) experiment is documented as an image displaying an area of a mouse (from a particular TS) in which some subsections of the mouse are highly coloured, as depicted in Figure 1. Areas of colour indicate that the gene is expressed in that location. Furthermore, the image provides some indication of the level (strength) of expression: the more intense the colour, the stronger the expression. Results are analysed manually under a microscope. A human expert determines in which structures the gene is expressed, and at what level of expression. Strength information is described using natural language terms such as strong, moderate, weak or present. For example, the gene *Bmp4* is strongly expressed in the future brain from TS15. These statements, called *textual annotations*, are a triple featuring: a gene, a tissue/stage and a level of expression. EMAGE also captures the raw spatial information by mapping the results onto 3D models of the developmental mouse forming *spatial annotations*.

### Existing EMAGE visualizations

There are essentially two categories of gene expression query: start with a gene and look for locations or start with a location and look for genes. EMAGE provides the same layout of results for both classes of query:

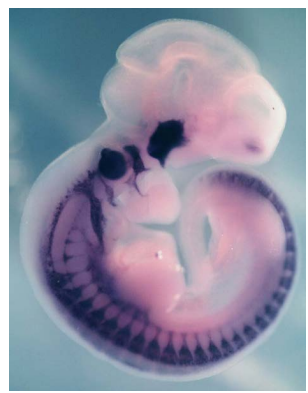

**Figure 1** A sample image of an experimental result from EMAGE (accession ID EMAGE:672). This image shows a mouse from TS17. The areas of colour show where the gene *Sox10* is expressed.

a table in which each row (e.g., Figure 2 (A)) provides a summary of an experimental procedure that obtained a relevant result. Additionally, EMAGE can summarise the results of many spatial annotations using a heat map; Figure 3 shows the location of gene expression in the *Wnt* Signalling Pathway at TS11. The third column provides all the individual gene expression spatial annotations, and the second column gathers these into a heat map with the colour intensity indicating the level of activity of the genes in a pathway.

GUDMAP [6] is a resources that focuses exclusively on the mouse's GenitoUrinary tract. Unlike EMAGE, it contains both Microarray and *in situ* experimental results. When a user query starts with a gene the results are returned in a table (an example row can be seen in Figure 2 (B)). Microarray results are reported using the standard heat map. *In situ* results are visualised in a bar chart: a positive bar indicates the gene is present, a negative bar shows that the gene is not expressed and a missing bar informs that there is no relevant information. This visualization is possible because GUDMAP is focusing on a small subset of the mouse's anatomy, it does not scale up to work with the entire anatomy and thus cannot be used within EMAGE.

When a GUDMAP query starts with an anatomical structure, the resultant table is very similar to that of EMAGE. A typical row is shown in Figure 2 (C).

Neither EMAGE nor GUDMAP provide a good visual mechanism that enables a user to interact with the database and explore the results. Furthermore, neither resource provides a technique for visualising the anatomy.

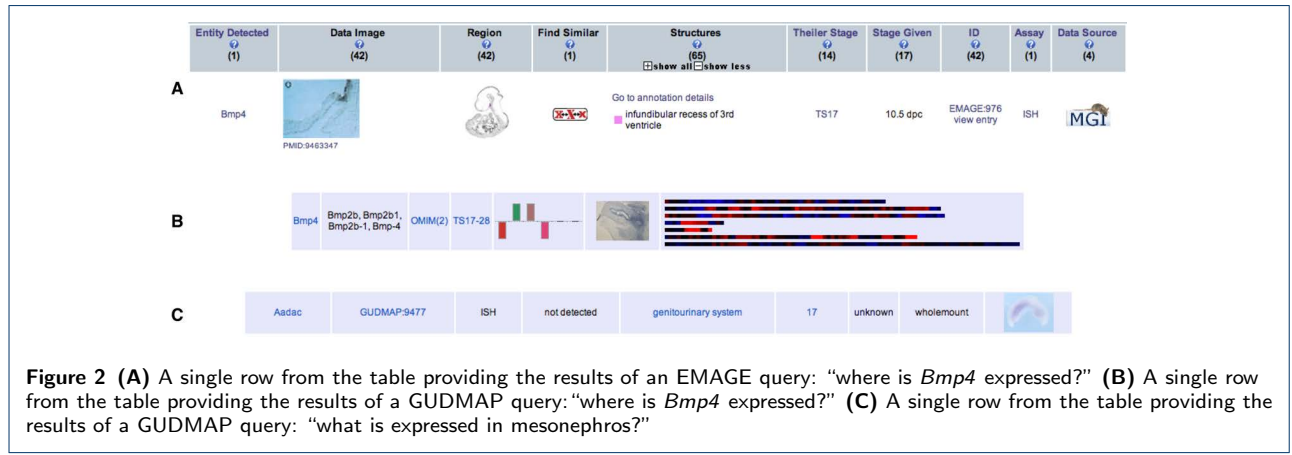

**Figure 2** **(A)** A single row from the table providing the results of an EMAGE query: "where is *Bmp4* expressed?" **(B)** A single row from the table providing the results of a GUDMAP query: "where is *Bmp4* expressed?" **(C)** A single row from the table providing the results of a GUDMAP query: "what is expressed in mesonephros?"

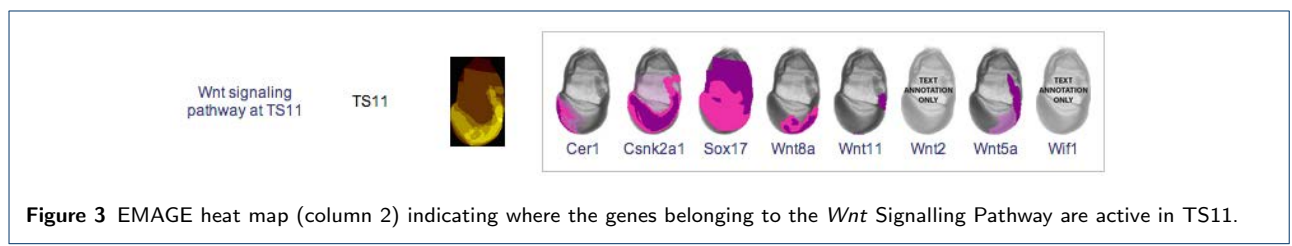

**Figure 3** EMAGE heat map (column 2) indicating where the genes belonging to the *Wnt* Signalling Pathway are active in TS11.

## Visualizations of gene expression data

The development of specialised visualizations for this use case was initially led by a focus group formed from the staff of EMAGE. This small team consisted of both biologists and bioinformaticians. They were responsible for making all the broad level (non implementation) decisions.

### Requirements

The first task for the focus group was to develop requirements for the tool being developed. Over the course of the project, these requirements evolved. Initially, there was one requirement: provide a visualization mechanism that allows the user to navigate the anatomy. Ultimately, this was extended and better defined, with the goal being to produce a visualization tool that would allow researchers to swiftly compare ontology-based *in situ* hybridization and/or immunohistochemistry profiles in three scenarios of increasing complexity:

1. To compare multiple gene expression profiles at a single stage of development in a single organism
2. To visualize the changing pattern of gene expression over time in a single organism
3. To visualize and cross-compare the changing pattern of gene expression over time in multiple organisms

The first and second scenario relate directly to the EMAGE database resource. The EMAGE database houses both text (ontology)-based annotation and spatial annotation and, whereas there are tools available on the EMAGE web interface to visualize spatial co-expression as a heat map, there is a need for a concomitant tool to allow researchers to visualize complex ontology-based annotation and explore coexpression in this way.

The third scenario is more complex and would allow a researcher to visualize and understand gene expression trajectories during development in different species such as mouse and human. In this respect, it is noteworthy that the HUDSEN resource [7] uses a similar spatial and ontology framework to EMAGE for annotating gene expression data, and a tool that would allow researchers to cross-compare *in situ* expression profiles across both database resources would be highly desirable. Due to time constraints this was left for future work.

There is also an important technical requirement. The latter stages of the mouse contain approximately 2000 anatomy structures. Additionally, up to 19000 genes can be expressed within a stage. The chosen visualization mechanism must scale appropriately.

### Choosing a visualization

With limited time available, it was necessary for the development team to restrict the choice of visualization(s) available to the focus group. Accordingly, the developers made the practical decision to develop a range of small prototypes using the d3.js framework [8]. This ensured that the focus group was not overwhelmed by the large range of plausible hierarchical visualizations [9]. Additionally, it meant that the developers did not need to develop a visualization completely.

The simple prototypes developed for this task focused on a couple of genes in only one Theiler Stage. The visualization mechanisms shown included those already in use: tree map and heat map. Mechanisms chosen to promote discussion: scatterplot, parallel coordinates and koalas to the max (see [10] for examples). The final group was populated by mechanisms that the development team felt would be most appropriate: force-directed graph, icicle, sunburst, cluster graph, and sankey diagram (see [10] for examples of these diagrams). During the focus group meeting, biological participants added two visualizations to the discussion: hive plots and circos diagrams.

One of the first decisions reached by the focus group was that they wanted to experiment rather than try and implement a visualization that someone else had already undertaken. This immediately ruled out the tree map and heat map. Other diagrams were disregarded for a variety of reasons: too complex (force-directed graphs and sankey), doubts about the ability of the mechanism to scale appropriately (circos and force-directed graphs), diagrams that EMAGE has already tried (force-directed graph and parallel coordinates) and finally, diagrams that did not seem appropriate (cluster graph, scatterplot and koalas to the max).

The diagrams that initially seemed most suitable for this exercise were the icicle, sunburst and hive plot. When forced to pick just one the focus group selected the sunburst. The hive plot was rejected because, after further consideration, the group decided it would not scale as well as the other two options. This left the sunburst and icicle; the sunburst won because the biological users felt it was the more interesting of the two.

### Generic sunbursts

A sunburst (e.g., see Figure 4) takes information organised within a tree structure, and displays the tree structure in a radial layout. Assuming the information is organised as a tree no organisational data is lost.

The centre of a sunburst diagram is the root node of the tree, with children of the root node being the first layer of blocks in the sunburst. Children sit directly around in the next layer of the sunburst, and so on, until the leaf nodes are reached at the edge of the diagram. The size and position of the blocks within the sunburst are used to indicate the structure, and organisation, of the data. Data attribute values are presented by colouring the nodes.

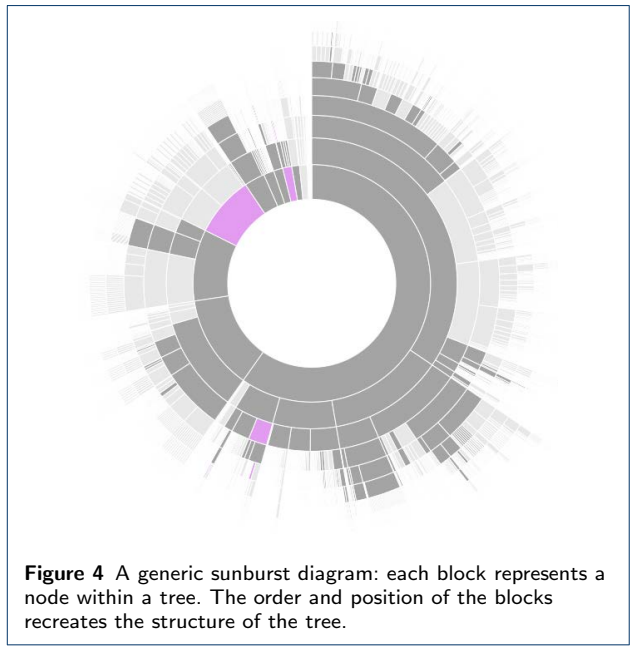

**Figure 4** A generic sunburst diagram: each block represents a node within a tree. The order and position of the blocks recreates the structure of the tree.

### Sunbursts for gene expression

Within this work the sunburst was used to represent the primary structure (tree) of the EMAP anatomy. Thus each node represents an anatomical structure apart from the root node, which is the mouse itself. The size of the nodes is directly related to the number of descendants they have. The larger the number, the greater the size of the node. The node's colour is used to present the level of expression for that node, the colour scheme chosen matches one already used by EMAGE (strong = red, moderate = yellow, weak= purple, not detected = cyan, propagated = pink and no expression information = grey).

Because the anatomy is a tree structure based on the *partOf* relationship, each structure (apart from the root node) is part of another structure. Therefore positive gene expression information can be propagated up the tree (e.g., see Figure 5). For example, if a gene is expressed in the digit, that gene is also expressed in the paw. Likewise, as the paw is part of the limb, the gene is expressed in the limb too. Propagated gene expression information is indicated through the application of a pink colour to a node. The colour pink is always chosen regardless of what level of positive expression the children have. The number of children with positive expression and the level of the expression in the children is not taken into account. Negative expression (not expressed) is not propagated.

The use of pink parent nodes to indicate propagated expression is particularly useful when the gene is expressed in a leaf node, as these nodes are often so small that they are hard to see. Colouring the ancestor nodes pink helps attract the user's attention to the leaf node.

In order to test the sunburst visualization, a simple user interface was developed (see Figure 5). When the user moves the mouse over a node, the box on the right of the screen is updated to show the name of the anatomical structure that node represents. If the node contains gene expression information it is displayed too.

The left hand box, in Figure 5, provides a simple menu system for refining the content of the sunburst. The user can select the gene(s) and ontology type (staged versus abstract). Moreover, the left box (Figure 5) provides a navigation function that allows the user to move from stage to stage. In this way the user can watch the expression profile change over time.

### Extending to icicles

Because it was simple to do, and the developer's favourite visualization, an icicle diagram was implemented too. The icicle (e.g., Figure 6) is essentially a linear sunburst, with the root node at the top and the children underneath. All other details are identical to the sunburst.

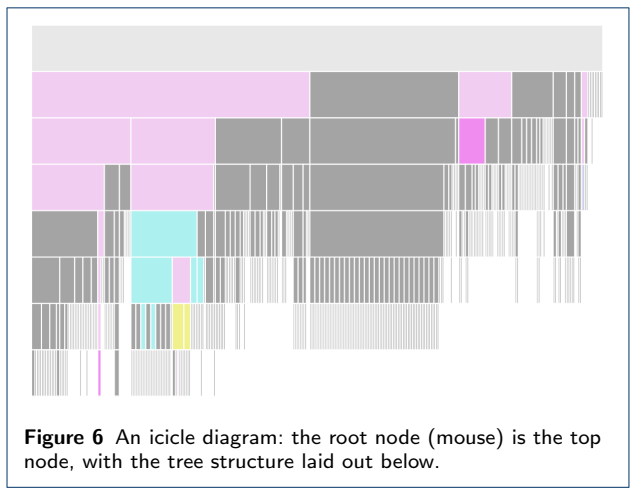

**Figure 6** An icicle diagram: the root node (mouse) is the top node, with the tree structure laid out below.

### Technical issues

The basic sunburst/icicle diagram was developed using the d3.js framework, with the surrounding application developed in Python using the Django framework [11]. These choices enabled rapid prototyping and interaction with the EMAGE focus group steering the work. Despite the many positives, there were a number of issues directly emanating from these decisions, in particular the selection of d3.js.

In order to apply the d3.js toolkit, the user must first generate a suitable tree structure and then ask d3.js to draw it. The speed of this depends upon two factors: the performance of the biologist's computer and the size of the tree. The latter was a particular problem.

As the mouse develops it becomes more complex. In the first Theiler Stage there are just 5 anatomical structures (including the mouse itself); by the final Theiler Stage the number of anatomical structures is approximately 2000. The more complex (i.e., more nodes) a tree has the longer it takes display it. At TS17 (1700 nodes) there is a noticeable delay in drawing the visualisations. Performance continues to decline until TS23 where upon it stabilises because stages 23 to 26 have a similar number of nodes. When comparing a sunburst with an identical icicle diagram, the sunburst takes longer to display due to the extra complexity of having to calculate and present the radial layout.

An additional issue is the failure of the zoom functionality on large sunbursts. By default, both the sunburst and icicle diagrams are equipped with a zoom. When the biologist clicks on a node, the parent node becomes the root of a new diagram in which the clicked node, and its siblings/children are increased in size. The zoom on the icicle diagram functioned perfectly, regardless of which stage it was asked to deal with. In contrast the sunburst zoom behaved erratically when asked to deal with any stage above TS14. Essentially, clicking on a node might produce the expected (i.e., "correct") sunburst or it might produce an incorrect sunburst which has a pie chart as the top node and a distorted set of children underneath. The confusion this caused among biological users forced the developers to switch off the zoom for the sunburst.

The problems with the sunburst zoom seem to be directly related to the number of nodes in the tree being rendered as stages with fewer than 500 nodes were unaffected. This appears to be a bug within the d3.js implementation; however, the developer did not respond to a request for help and thus we were unable to further analyse this issue.

## Building a gene expression query

Whilst it is possible to create simple gene expression queries via the main sunburst/icicle pages another, more powerful, option is provided. The so-called "advanced search page" provides a cloud of nodes (i.e., cluster graph [12]) that represents the textual annotations present at a particular stage (e.g. Figure 7). This was implemented, despite the focus group rejecting the use of cluster graphs, because the developers felt it would be an interesting experiment.

Figure 7 shows the cluster graph for TS12. Each node represents a gene involved in a textual annotation at this stage. The size of the node directly corresponds to the number of textual annotations that gene has at this stage, i.e., the greater the number of annotations the larger the node. When the mouse pointer hovers

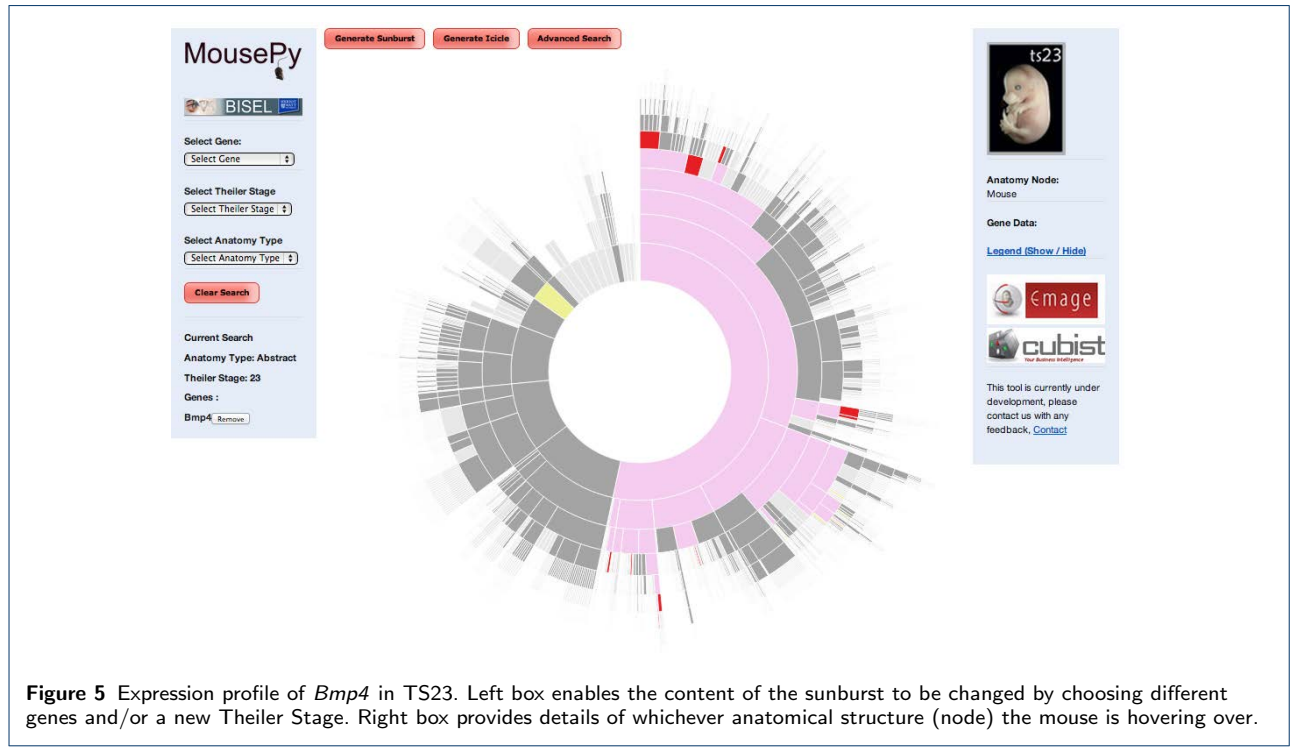

**Figure 5** Expression profile of *Bmp4* in TS23. Left box enables the content of the sunburst to be changed by choosing different genes and/or a new Theiler Stage. Right box provides details of whichever anatomical structure (node) the mouse is hovering over.

over a node, a tool tip reports the number textual annotations contained within the database.

To make them easier to find in large clouds, genes (nodes) are clustered alphabetically meaning that the members of a gene family (e.g., *Bmp2*, *Bmp3* ... *Bmp10*) are adjacent to one another. If there are too many nodes on screen, it is possible to filter the genes by selecting a subcomponent of the anatomy. For example, rather than look at a cloud of almost 19000 nodes for the whole mouse at TS23, the user can choose to focus on a particular structure (e.g., the heart) and thus considerably reduce the number of nodes (genes) visible. To illustrate, the cloud in Figure 7 is reduced to just three nodes when the user focuses upon the "eye".

Clicking on a node turns it red, and adds that gene to the query the user is building. The Theiler Stage can be changed by using a listbox (not shown). In this manner the user can ask, *"where are the following genes expressed in my chosen stage?"* When the user is happy with his/her query, (s)he can generate either a sunburst or an icicle diagram.

## Evaluation

A short evaluation was undertaken using staff members from EMAGE as test users. During the evaluation test subjects were given a tutorial that guided them through the prototype's main features, subsequently they were asked to complete a range of SUS questionnaires (one each for the icicle, sunburst and advanced search page), a QUIS questionnaire for the complete prototype and a questionnaire that asked for their favoured visualization (sunburst versus icicle).

Whilst the evaluation has been too small to draw statistically valid conclusions the following highlights are interesting. Firstly, the sunburst seems the most popular visualization with users stating a preference for the radial layout. Nevertheless, the icicle achieves higher SUS scores for usability. Secondly, the abstract anatomy was considered far superior to the staged anatomy (see the Discussion section for more information). Thirdly, although it was something of an afterthought, the advanced search page was the most successful component with the highest SUS scores and the least reported issues.

## Discussion

### Ontological discussion

As discussed previously, the sunburst/icicle diagrams are ideal for presenting trees; however, if the structure is a graph they lose information. The current version of the EMAP anatomy is a DAG; however, this evolved from an earlier version in which the structure was a tree. Today, the EMAGE community still view the tree as the primary representation of the anatomy. Accordingly, it is the tree that is used for the sunburst/icicle depictions not the larger DAG. Whilst the sunburst/icicle contains less information than the full anatomy, it is based on the biologist's preferences.

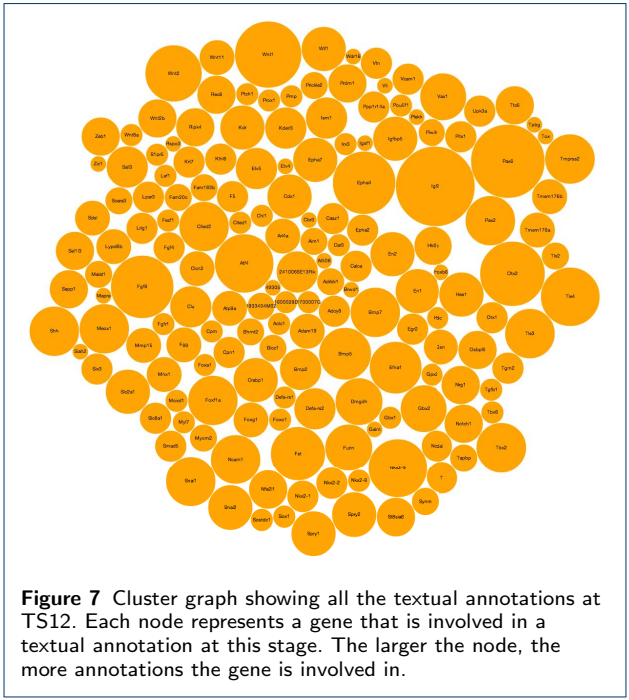

**Figure 7** Cluster graph showing all the textual annotations at TS12. Each node represents a gene that is involved in a textual annotation at this stage. The larger the node, the more annotations the gene is involved in.

Another important feature of the EMAP anatomy is the so-called "abstract anatomy". Within EMAP there is one anatomy for each stage (so-called "staged anatomy") and one anatomy that aggregates the stages into one structure (i.e., rather than having one heart for each stage, the abstract anatomy has one heart that covers every stage at which the heart exists). All of the screenshots so far have shown the sunburst/icicle with the staged anatomy; however, usability tests have revealed that this is not the preferred option for the biologists. Figure 8 demonstrates why. As the mouse develops over time anatomical structures appear and disappear, this causes the corresponding nodes to appear and disappear too. Ultimately, this effects the layout of the visualization resulting in structures "moving" around the diagram. As nodes (anatomical structures) do not have a fixed location, this makes it difficult for a biologist to quickly scan the diagram. It would be far simpler if the location of the anatomical structures (nodes) remained constant. This can be achieved through the use of the abstract anatomy. Figure 9 shows the same information as Figure 8, but this time using the abstract anatomy. In this sunburst, light grey nodes represent anatomical structures that do not exist at the current developmental stage, grey nodes represent structures that exist but do not have any expression information, and the previously described colour scheme depicts nodes with associated gene expression information. Thus the change between stages is represented by the change of node colour rather than the change of the basic layout of the diagram.

Whilst the above text focuses upon the sunburst diagram, the comments are equally true for the icicle.

Biological relevance

For the biologists on the focus group, this work represented an opportunity to expand the range of interaction mechanisms offered by EMAGE. The original goal was to create a prototype that would allow the biological end user to navigate through the resource, exploring the original EMAGE web pages if desired. Whilst the tool does not currently link into the original EMAGE resource, it would be easy to extend it in such a way that those links exist. For example, when the biologist hovers his/her mouse over a node, it would be possible to provide a hoverbox containing a description of the textual annotation documented by the node and a link to the original EMAGE web page describing the source experiment.

As development of the application progressed, it became clear that the focus group found the sunburst visualization to be more useful and visually appealing than the icicle. The icicle is a long diagram, which means that it looks good on widescreen monitors. The space saving radial design of the sunburst often leaves large areas of whitespace that some end users objected to. However, the spacing saving design of the sunburst was its greatest strength. Because it is so small, multiple sunbursts can be placed on the same screen. This provides the ability to compare and contrast expression information for a set of genes. Such functionality can be used to compare the expression profile of the same gene over time. It was deemed especially useful when considering the expression profile of a set of genes found in a particular pathway or a family of genes.

Figure 10 shows the *Bmp* family (*Bmp2* ... *Bmp10*) at TS17. In one glance it is clear that the expression profiles for *Bmp2*, *Bmp6* and *Bmp8* are all contained within *Bmp4*; i.e., the list of structures in which *Bmp2* is expressed is a subset of those in which *Bmp4* is expressed, likewise *Bmp6* - *Bmp4* and *Bmp8* - *Bmp4*. *Bmp10* is distinct because it is expressed in the cardiovascular system and *Bmp4* is not. This example demonstrates both the requirement to learn the basic layout of the sunburst and the benefits of doing so. The level two nodes represent the larger systems: cardiovascular, nervous etc., and by learning their position within the sunburst it is possible to quickly survey large amounts of information. If the sunbursts were then linked to the underlying EMAGE web pages, as described above, this would provide an efficient way of surveying and then "deep diving" into the data.

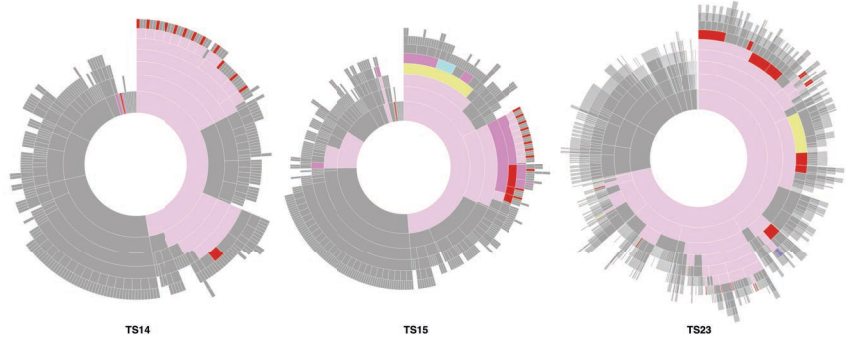

**Figure 8** Depicting the expression profile of the gene *Shh* across time using the staged anatomy. Notice that the the shape and size of the sunburst changes from stage to stage as the constituent tissues change.

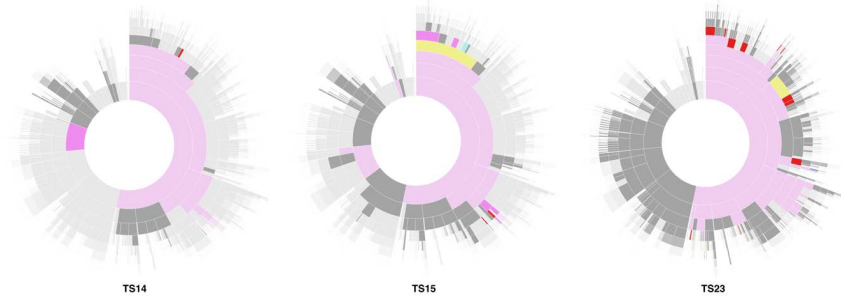

**Figure 9** Sunbursts with the abstract anatomy: the same information as Figure 8, but this time displayed using the abstract anatomy. Notice that unlike Figure 8, the shape and size of the sunbursts remain constant over time. Instead of changing the nodes, the colour of the nodes indicate whether or not a tissue is present at this stage.

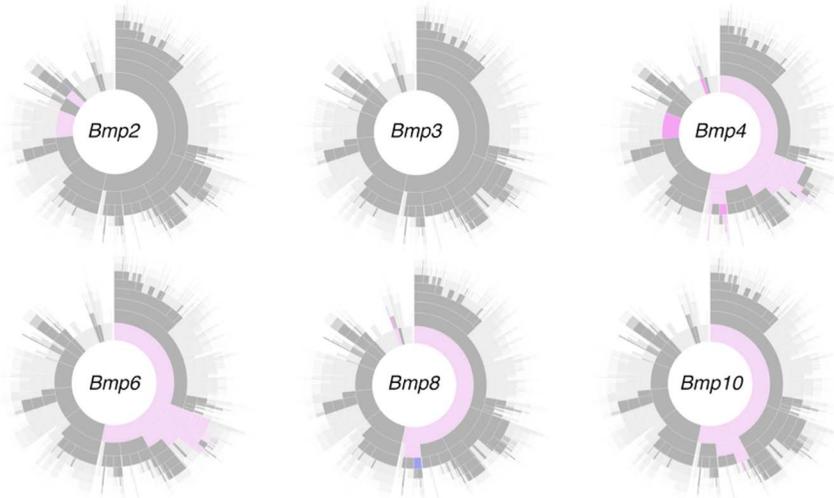

**Figure 10** Sunbursts using the abstract anatomy showing the expression profile for genes from the *Bmp* family at TS17.

Perhaps the best indication of the value of this work is the fact that the EMAGE developers are keen to take over the effort described in this paper, and include the visualizations within their web presence.

Labelling issues

One of the most glaring omissions from the current system, is the lack of labels for the anatomy structures (nodes in the sunburst or icicle). However, this is a more complex issue than first appears.

Initially, it was assumed that every node should have a label. Yet experimentation with the focus group suggested that only the major systems (e.g., cardiovascular) need to be labelled. The other nodes can be identified with the use of a mouseover or hoverbox. Whilst this may seem strange, it enables the diagrams to function as an overview and thus supports the "survey and deep dive" behaviour described previously.

The second presumed difficulty was that the labels would be too big for the nodes. Again, the focus group provided a solution. They suggested a list of common abbreviations that could be used instead of the full name, e.g., CNS instead of Central Nervous System.

The remaining difficulty relates exclusively to the sunburst: how to write the labels radially so that they fit into the nodes of the sunburst? This issue was not settled to the satisfaction of the focus group within the project and remains as further work.

Perceived issues with the Cluster Graph

Within the prototype, a cluster graph enables users to find genes they wish to see the gene expression profiles for. It was assumed that this visualization would encounter a number of issues. In particular, it was suggested that the user may struggle to see the nodes in some stages: TS23 has 19000 genes and thus 19000 nodes making each node very small. Moreover, despite the genes being clustered alphabetically, there was a feeling that it may be difficult to find the desired gene.

During the evaluation neither of these issued concerned the biological users; the test subjects enjoyed the ability to see all the textual annotations on a single screen. The ability to search by gene symbol or filter by anatomical location seemed to resolve the above issues. For future development it is intended to enable filtering by genes and gene co-expression across Theiler stages.

Related work

Increasingly data-sets within the life sciences are approaching sizes which are not manageable by humans and as such usable visualizations are vital in helping human researchers navigate this data [13]. The analysis of gene expression data in particular is a complex task for human researchers. Within the current use case there are approximately 19000 genes and 2000 anatomy structures leading to a large and combinatorially complex dataset. Wang *et. al* propose that interactive visualization techniques, such as those used within this work, are a good way of navigating complex big data [14].

Accordingly, the life sciences are a very active area for visualization. For example heat maps have been used to demonstrate Anisotropic Flocking Behaviour [15], hive plots [16] have been used to visualise gene regulatory networks. A variety of partition graphs [17] such as sunbursts [18], icicles [19], and partition diagrams [20] have been used to visualise ancestry. Dendrograms, e.g. [21], are the visualizations traditionally used to represent differences between species.

Circos [22] was designed for the visualization of related data using a circular layout, this approach has been widely used within bioinformatics and other fields. This type of circular layout has been found to be hard to interpret [16].

Force directed graphs [23] are a commonly used means of addressing the problem of efficiently displaying a network layout, e.g., Cytoscape [24] is a toolkit for generating biological graphs. Hive plots have been advanced as an alternative to circular type layouts (e.g. circos) and force directed graphs. Hive plots have been claimed to enable better comprehension of the relationships within a dataset. Hive plots provide a rational and transparent visualization technique for complex network types by laying out nodes on several radially oriented linear axes with a co-ordinate system based on a node's structural properties. Hive plots have been found to be useful ways of visualising several types of complex networks [25, 26, 27].

In this work the goal is to visualise gene expression data within the context of the anatomy. We are currently unaware of any attempts to visually present gene expression data for the mouse within the context of the anatomy that does not feature a picture of the mouse itself. Whilst this is a valid approach, we believe a BI style abstraction provides a complementary mechanism. In particular, it provides the ability to navigate and browse the data in a simple, standardised way.

Experiments have shown that the standardisation is very important to our potential user group. We have achieved a standardised presentation by using the sunburst/icicle with the abstract anatomy. It is this ability that separates the sunburst/icicle from the other visualizations explored. The additional complexity of certain diagrams, and the need to re-learn a new layout with every change to the data presented, seemed to irate some potential users. There is a scalability issue too. Some of the stages have nearly 2000 anatomical nodes and it is questionable how usable, for example, a circos diagram of that size would be.

## Conclusions

This work has focused on the application of sunburst and icicle diagrams in order to represent both the de-velopmental mouse anatomy, and the expression profile of genes within the context of that anatomy. By applying the so-called abstract anatomy we were able to develop a standardised layout that did not change between developmental stages. Each node represents a single anatomical structure, and that node remains fixed in both location and size regardless of the stage viewed.

In partnership with biologists from the use case, a prototype implementation was developed and tested. The prototype enables the search and navigation of a large dataset which is combinatorially explosive, from user evaluations it is apparent that biologists find these methods of visualization more useful than existing methods. Whilst the icicle diagram provided a good way to browse through the data on a large widescreen monitor, the sunburst was ideal for smaller screens. Additionally, the sunburst provides a very good visual overview of where a gene is expressed at a particular Theiler Stage. The space saving quality of the sunburst allows multiple sunbursts to be displayed on the screen at once. This enables multiple genes, or stages, to be compared in one glance: a feature the biological test users found to be very useful.

Work is currently ongoing to incorporate this work into the EMAGE user experience.

**List of abbreviations used**
BI: *Business Intelligence*, DAG: *Direct Acyclic Graph*, EMAGE: *eMouse Atlas of Gene Expression*, EMAP: *developmental mouse anatomy*, SBI: *Semantic Business Intelligence*, TS: *Theiler Stage*.

**Competing interests**
The authors declare that they have no competing interests.

**Author's Contributions**
KCM wrote the first draft of the manuscript, and the majority of the second draft. CA added the user requirements. AT produced the related work. AB and RB edited the document and contributed extensions to the second draft.

**Acknowledgements**
This work is part of the CUBIST project (Combining and Uniting Business Intelligence with Semantic Technologies), funded by the European Commission's 7th Framework Programme of ICT under topic 4.3: 'Intelligent Information Management'.
The authors are thankful for the advice and guidance provided by members of the EMAGE team.

**Author details**
[1]Heriot-Watt University, EH14 4AS Edinburgh, UK. [2]MRC Human Genetics Unit, University of Edinburgh, Edinburgh, UK.